\documentclass[journal]{IEEEtran}

\usepackage{amsmath,amssymb}
\usepackage{graphicx}
\newcommand{\Tr}{\mathrm{Tr}}
\newcommand{\sumhight}{{\displaystyle\vphantom{\sum_{i=1}^m}}}

\newcommand{\bra}[1]{\langle #1 |}
\newcommand{\ket}[1]{| #1 \rangle}
\newcommand{\bracket}[2]{\langle #1 | #2 \rangle}

\newcommand{\thmbuffer}{\vspace{6pt}}

\newcommand{\eff}{{\rm eff}}
\newcommand{\opt}{{\rm opt}}
\newcommand{\rank}{{\rm rank}}
\renewcommand{\det}{\text{det }}
\newcommand{\st}{\text{s.t.}}
\newcommand{\mcH}{\mathcal{H}}
\newcommand{\mcR}{\mathcal{R}}
\newcommand{\mcM}{\mathcal{M}}
\newcommand{\mcI}{\mathcal{I}}

\newtheorem{theorem}{Theorem}
\newtheorem{lemma}{Lemma}
\newtheorem{corollary}{Corollary}[theorem]

\begin{document}
%
\title{Optimal Encoding of Classical Information in a Quantum Medium}
\author{Noam~Elron 
        and~Yonina~C.~Eldar,~\IEEEmembership{Member,~IEEE}
\thanks{Manuscript received January 66, 2666; revised November 66, 2666.
        }
\thanks{The authors are with the Technion - Israel Institute of Technology, Haifa
        32000, Israel (e-mail: nelron@tx.technion.ac.il; yonina@ee.technion.ac.il).}}

\markboth{Submitted to IEEE TRANSACTIONS ON INFORMATION THEORY}
         {Elron & Eldar: Optimal Encoding of Classical Information in a Quantum Medium}


\maketitle

\begin{abstract}
We investigate optimal encoding and retrieval of digital data, when
the storage/communication medium is described by quantum mechanics.
We assume an $m$-ary alphabet with arbitrary prior distribution,
and an $n$-dimensional quantum system.
Under these constraints, we seek an encoding-retrieval setup,
comprised of code-states and a quantum measurement, which maximizes
the probability of correct detection.
In our development, we consider two cases.
In the first, the measurement is predefined and we seek the optimal
code-states.
In the second, optimization is performed on both the code-states and
the measurement.

We show that one cannot outperform `pseudo-classical transmission', in
which we transmit $n$ symbols with orthogonal code-states, and discard
the remaining symbols.
However, such pseudo-classical transmission is not the only optimum.
We fully characterize the collection of optimal setups, and briefly discuss
the links between our findings and applications such as quantum key distribution
and quantum computing.
We conclude with a number of results concerning the design under an alternative
optimality criterion, the worst-case posterior probability, which serves as a
measure of the retrieval reliability.
\end{abstract}

\begin{keywords}
transmitter design, quantum detection, quantum key distribution,
semidefinite programming, bilinear matrix inequality.
\end{keywords}

%
\IEEEpeerreviewmaketitle

\section{Introduction}

\PARstart{U}{nderlying} any scheme for the storage or transmission of
information is a physical medium.
The encoding and the retrieval of information must therefore involve
considerations as to the nature of the medium, with regard to possible
corruption of the retrieved data, due to interaction with the environment
or to physical limitations of the medium itself.
Examples of media and information encoding range from letters printed
in ink on paper, through electric charge stored in a capacitor, to
photons travelling through an optical fiber.
This work is concerned with the encoding of digital information in media,
whose physics is described by the laws of quantum
mechanics \cite{Peres:BOOK}.

We concentrate on digital information with a finite alphabet, i.e. the
data is one of $m$ possible messages, each one associated with a prior
probability $p_i$.
Retrieval of the data is done by performing a measurement, thereby
detecting the state of the system.

There are several common criteria for the assessment of information
retrieval, which can, for the most part, be divided into two categories.
The first is comprised of criteria whose motivation stems from information
theory (e.g. mutual information \cite{CoverThomas}).
The second type of criteria aim to measure the reliability of the
``per symbol'' retrieval, without taking into account any pre- or
post-processing (channel coding).
The criteria we address in this work are of the second kind.

The laws of quantum mechanics state that the outcome of a
measurement, in our case an attempt to retrieve the encoded
symbol, is random.
Thus, a quantum encoding-retrieval setup is characterized by
the transition probabilities
\[
\Pr \{i|j\} \triangleq \Pr\{\text{out = symbol $i$}|\text{in = symbol $j$}\}.
\]
This is reminiscent of the more common classical setups, but whereas
the randomness there is induced by noise from the environment,
in the quantum case, the randomness is inherent in the system itself.

The state of a quantum system is mathematically represented by
a unit trace positive semidefinite operator $\rho$ on an
$n$-dimensional Hilbert space $\mcH$.
Encoding digital information in a quantum system is done by preparing
the system in one of $m$ predefined states $\{\rho_i\}_{i=1}^m$, each
associated with one of the possible messages.
Retrieval is achieved by performing a measurement, and determining in
which of these predetermined states the system has been prepared.

However, if a quantum system is in one of several states whose range
spaces are not orthogonal, i.e. $\rho_i\rho_j\neq0$,
then no measurement permitted in quantum mechanics can determine
without fail which of the states is present; there is a non-zero
probability of detection error, i.e. $\Pr\{i|j\}>0$ for $i \neq j$.
The question is then, what valid quantum measurement would yield favorable
detection performance.

A popular measure of performance, and the one which is the
main interest of this work, is the probability of correct detection
\[
P_d = \sum_{i=1}^m p_i \Pr \{i|i\}.
\]
One of the contributions of this work is a complete characterization
of the encoding-retrieval setups which maximize $P_d$ under the
constraints imposed by the postulates of quantum mechanics.
We also present several new results concerning a different performance
measure, the worst-case posterior probability, which is defined in
Section \ref{posterior}.

The focus of this paper is the design of a complete digital communications
channel (or memory unit), in which the designer can choose both the
code-states $\rho_i$ and the detection measurement.
We assume that the nature of the data, which is designated by the number of
possible symbols $m$ and their prior probabilities $p_i$, is known.
We also assume that the dimension $n$ of the quantum system is given.
The dimension of the quantum system determines the ability of the medium to
transmit (or store) data reliably, much like the signal to noise ratio in
classical systems.

Thus, we seek the optimal setup, comprised of code-states and a measurement,
that maximize $P_d$ under a constraint of the dimension $n$ of the system.
We find the maximum attainable value of $P_d$ for data with an arbitrary prior
distribution, and completely characterize the optimal setups, which achieve this
value of $P_d$.

When the number of symbols $m$ is no larger than the dimension of the Hilbert
space $n$, one can simply choose $\rho_i$ as orthogonal pure states and attain
perfect detection.
When $m>n$ this is no longer possible, and quantum encoding becomes non-trivial.

Motivation for using many symbols in a quantum system of low dimension
may stem from benefits, which a protocol provides, for which one is
willing to sacrifice the probability of detection or the information
rate.
For instance, in protocols of quantum key distribution \cite{QKDtutorial}
the use of many states enables the detection of eavesdropping on the
communication.
In Section \ref{QKD} we elaborate on this point.
Another possible scenario is a quantum computation, which has a finite
number of possible outputs $m$, and where for reasons of implementation
complexity one cannot create a system large enough (with enough qubits
such that $m \leq n$).

The problem of distinguishing among a collection of \emph{specified}
quantum states, i.e. when the code-states $\rho_i$ are a given, is
regularly referred to as \emph{quantum detection} or \emph{quantum
state discrimination}, and has been studied in detail.
Necessary and sufficient conditions for an optimal measurement,
which maximizes the probability of correct detection $P_d$,
have been derived \cite{Holevo1,YKL1,YoninaMegretskiVerghese1}.
Explicit solutions to the problem are known in some particular cases
\cite{Helstrom1,CBH,OBH,BKMH,YoninaForney1}, including ensembles
obeying a large class of symmetries \cite{YoninaMegretskiVerghese2}.
The optimal measurement can also be calculated numerically, to within
arbitrary accuracy, and in polynomial complexity \cite{YoninaMegretskiVerghese1}.

Several alternative approaches have also been investigated.
These include optimization with regard to other performance criteria,
such as mutual information \cite{Holevo1} or the worst-case posterior
probability \cite{KWER}.
Another approach is \emph{unambiguous detection} \cite{Ivanovic,YoninaUnambig,YoninaStojnicHassibi}
in which one allows for an inconclusive result but does not allow for error.
More recently, interest has grown in detection in a noisy environment
\cite{SaMoHi,VilnrotterLau2,ConchaPoor}, and in situations where the
states are only partially known \cite{NoamYonina1} or the prior probabilities
not specified \cite{DaSK}.

In Section \ref{formulation}, the problem is presented in more detail.
Then, in Section \ref{codestates}, we show that the optimal code-states for
a predetermined measurement are states which lie in the eigenspaces of the
measurement operators associated with the maximal eigenvalues.
This result is of interest both in its own right, and as part of the design
of complete optimal encoding-retrieval setups.

Sections \ref{maxPd} and \ref{Characterization} are the heart of this work.
In Section \ref{maxPd} we show that when encoding digital information in a
quantum system of dimension $n$, the maximum attainable probability of
correct detection may be achieved by simply discarding $m-n$ of the symbols
and using an orthonormal set to encode the remaining $n$ symbols with
perfect reconstruction.
We dub this method \emph{pseudo-classical} transmission.
This is, however, not the only possible encoding-retrieval setup which
achieves the maximal value of $P_d$.
In Section \ref{Characterization} we show that all setups that attain the
maximum are composed of pure code-states and of rank-1 measurement operators,
and fully characterize the collection of optimal setups.
The importance of finding all the optimal setups is discussed in Subsection
\ref{QKD}, where we outline possible use of our results in the analysis
of quantum communication and computation protocols.

In Section \ref{posterior} we explore performance in relation to a measure
of the reliability of the outcome.
We introduce the \emph{worst-case posterior probability}, denoted $P_p$.
Again, when $m>n$ and perfect communication is impossible, the output can
never be fully reliable.
We provide a simple method for finding an upper bound on $P_p$ for arbitrary
states $\rho_i$ and prior probabilities $p_i$.

Regrettably, for a large family of encoding-retrieval setups, $P_p$
is ill-defined.
For this reason, we also define a variation on $P_p$ that we name the
\emph{effective} worst-case posterior probability.
We investigate how one should choose the code-states, which represent
discarded symbols in pseudo-classical transmission, in order to increase
the reliability of the output, while still attaining maximal $P_d$.
We develop an upper bound on $P_p^{\eff}$, and present a choice which
attains it.

\section{Problem Formulation}
\label{formulation}

\subsection{Notation}

According to the postulates of quantum mechanics \cite{Peres:BOOK},
a physical system is mathematically represented by an $n$-dimensional
complex Hilbert space $\mcH$.
The state of the system $\rho$ is represented by a positive semidefinite
(PSD) Hermitian operator on $\mcH$, such that $\Tr(\rho)=1$.
Throughout, we shall use the notation $A\geq0 $ to indicate
that an operator $A$ is PSD, and the notation $A \geq B$ to
imply that $A-B$ is PSD.
If $\rank(\rho)=1$, then it is known as a pure state.

As is customary in work relating to quantum theory, we shall
use Dirac's notation of linear algebra, wherein a vector is
denoted by $\ket{u}$, its Hermitian conjugate by $\bra{u}$,
and inner and outer products are signified by $\bracket{u}{v}$
and $\ket{u}\bra{v}$ respectively.
We do not assume that $\ket{u}$ is normalized.
We denote by $\mcR(A)$ the range space of a Hermitian operator
$A$, and by $\mcM(A)$ the eigenspace of its maximal eigenvalue.

\subsection{Encoding Data in Quantum Media}

We wish to encode digital information in a quantum medium.
The information is represented by an $m$-ary alphabet,
where each symbol has a prior probability $p_i$.
Without loss of generality, we assume that the prior
distribution obeys $p_1 \geq p_2 \geq \cdots \geq p_m > 0$.
The encoding is achieved by associating with each symbol a
predefined quantum state $\rho_i$, and preparing the system
in the appropriate state.
We shall refer to the states $\rho_i$ as \emph{code-states}.
To a set of code-states $\{\rho_i\}_{i=1}^m$ we refer as an
\emph{ensemble}.
Whenever an ensemble is arbitrary, we assume that it spans\footnote{If
it does not span $\mcH$, the problem can always be projected
onto the subspace which it spans.} $\mcH$.

Retrieval of the information is accomplished by using a
positive operator valued measurement (POVM), which is a set
of $m$ operators $\Pi=\{\Pi_i\}_{i=1}^m$, which satisfy
\begin{align}
\Pi_i &\geq 0,\qquad 1 \leq i \leq m \nonumber \\
\sum_{i=1}^m \Pi_i &= I. \nonumber
\end{align}
This is the most general type of measurement allowed by the
laws of quantum physics.

The measurement results in one of $m$ possible outcomes,
where, given that the state of the system is $\rho$, the
probability of the $i$-th outcome is
\[
\Pr \{ i \} = \Tr ( \Pi_i \rho ).
\]
Thus, the probability of correctly detecting the encoded
message is
\[
P_d = P_d(\Pi_i,\rho_i) = \sum_{i=1}^m p_i \Tr (\Pi_i \rho_i).
\]
In this work we use $P_d$ as the main criterion for measuring
the quality of an encoding-retrieval setup.

In the next section we find the optimal code-states, in
the sense of maximal $P_d$, for a given measurement.
We then characterize, in Sections \ref{maxPd} and \ref{Characterization},
all optimal encoding-retrieval setups, when the design
specifications are the nature of the data (the prior
probabilities $p_i$), and the dimension $n$ of the quantum
system.

In Section \ref{posterior} we develop several results concerning
an alternative measure of performance, the \emph{worst-case
posterior probability}.
This criterion is an indicator of the reliability of the output,
and is defined at the beginning of Section \ref{posterior}.

\section{Designing Code-States for an Arbitrary Measurement}
\label{codestates}

In this section we answer the following question.
If the detector, i.e. the measurement $\Pi$, and the prior
probabilities of the data $p_i$ are predetermined, what would
be a good choice of code-states $\rho_i$ to encode the data
in a quantum medium of dimension $n$, in terms of $P_d$?
This question is of interest, due to possible implementation
restrictions on the detector.
As indicated in the introduction, the reverse situation, that
of designing a measurement to discriminate among arbitrary states,
has been thoroughly studied.

Our result is stated formally in Theorem \ref{thm1}.

\thmbuffer
\begin{theorem} \label{thm1}
Let $\{p_i\}_{i=1}^m$ be a probability distribution, and let
$\{\Pi_i\}_{i=1}^m$ be the measurement operators of a detector.
An ensemble of quantum states $\{\rho_i\}_{i=1}^m$ maximizes $P_d$
if and only if
\[
\mcR(\rho_i) \subseteq \mcM(\Pi_i).
\]
Denoting the maximal eigenvalue of $\Pi_i$ as $\sigma_{\Pi_i}^{\max}$, the maximal
probability of correct detection is given by
\[
P_d^{\opt} = \sum_{i=1}^m p_i \sigma_{\Pi_i}^{\max}.
\]
\end{theorem}
\thmbuffer

Note that for all $i$ such that $\Pi_i=0$, one has that $\mcM(\Pi)=\mcH$, and any
choice of $\rho_i$ is optimal.

\begin{proof}
The optimal states $\hat{\rho}_i$ are a solution to
\begin{align}
& \max_{\rho_i} \sum_{i=1}^m p_i \Tr (\Pi_i \rho_i) \label{firstProb} \\
& \qquad \st \begin{cases}
            \rho_i \geq 0, \\
            \Tr(\rho_i) = 1. 
                     \end{cases} \nonumber
\end{align}

The objective function in \eqref{firstProb} is additive in the
variables $\rho_i$, and the constraints on each of the $\rho_i$ are
independent.
Hence, 
\eqref{firstProb} is separable in $i$, i.e. the
states $\hat{\rho}_i$ are optimal if and only if they are also the
solutions to $m$ problems of the form (one for each $i$)
\begin{align}
& \max_{\rho} \Tr (\Pi \rho) \label{sepProb} \\
& \qquad \st \begin{cases}
            \rho \geq 0, \\
            \Tr(\rho) = 1. 
                     \end{cases} \nonumber
\end{align}

Any quantum state $\rho$, such that $\rho\geq0$ and $\Tr(\rho)=1$,
has an eigendecomposition of the form
\[
\rho = \sum_{j=1}^n g_j \ket{u_j}\bra{u_j},
\]
where $g_{j}\geq0$, $\sum_{j=1}^n g_{j}=1$, and $\bracket{u_j}{u_j}=1$.
Since $\Pi\geq0$, we have that
\begin{align}
\Tr(\Pi \rho) &= \sum_{j=1}^n g_j \bra{u_j}\Pi\ket{u_j} \nonumber \\
              &\leq \bra{\hat{u}}\Pi\ket{\hat{u}} \sum_{j=1}^n g_j \nonumber \\
              &= \bra{\hat{u}}\Pi\ket{\hat{u}} \nonumber \\
              &\leq \sigma_{\Pi}^{\max}, \nonumber
\end{align}
where $\bra{\hat{u}}\Pi\ket{\hat{u}} = \max_j \bra{u_j}\Pi\ket{u_j}$,
and $\sigma_{\Pi}^{\max}$ is the largest eigenvalue of $\Pi$.
If $\Pi=0$ then the upper bound is zero and any $\rho\geq0$ is optimal.
When $\Pi\neq0$, equality is achieved if $\Tr(\Pi\rho)=\sigma_{\Pi}^{\max}$,
i.e. \emph{only} when $\rho$ lies in the eigenspace corresponding
to $\sigma_{\Pi}^{\max}$.
\end{proof}
\thmbuffer

Note that the optimal code-states $\hat{\rho}_i$ are independent of
each other and of the prior probabilities $p_i$.
Also note that the optima (the solutions of the problem \eqref{firstProb})
form a convex set.

\thmbuffer
\begin{corollary} \label{cor1}
If for all $i$, $\dim \mcM(\Pi_i) = 1$, then the ensemble which
maximizes $P_d$ is unique.
\end{corollary}
\thmbuffer

\begin{proof}
When $\dim \mcM(\Pi_i) = 1$ then $\rho_i$ must be the pure state
which spans $\mcM(\Pi_i)$, and which is unique (due to the
requirement of normalization).
If this is true for all $i$, then the entire set of code-states
is unique.
\end{proof}
\thmbuffer

In applications, one may have the freedom to choose which symbol
will be detected by which of the detection operators.
Recalling that we assumed the prior probabilities $p_i$ to be sorted
in descending order, maximal $P_d$ can be attained when the detection
operators are sorted such that
$\sigma_{\Pi_1}^{\max}\geq\sigma_{\Pi_2}^{\max}\geq\cdots\geq\sigma_{\Pi_m}^{\max}$.
Doing this, and selecting the optimal code-states as above, would lead
to the maximal value of $P_d=\sum_i p_i \sigma_{\Pi_i}^{\max}$.

\section{Optimal Quantum Encoding}
\label{maxPd}

We now find the maximal attainable value of $P_d$ when encoding data
in a quantum medium.
We assume that the nature of the data itself, which is manifested in
the prior probabilities $p_i$, is predetermined, and so is the quantum
system itself (i.e. the dimension $n$).
We aim to find an 
encoding-retrieval setup
that maximizes $P_d$.

Thus, our goal is to find the solutions to
\begin{align}
& \max_{\Pi_i,\rho_i} \sum_{i=1}^m p_i \Tr (\Pi_i \rho_i) \label{origProb} \\
& \qquad \st \begin{cases}
            \sumhight\rho_i \geq 0, \;\qquad \Tr(\rho_i) = 1, \\ 
            \Pi_i \geq 0, \qquad \displaystyle\sum_{i=1}^m \Pi_i = I. 
                     \end{cases} \nonumber
\end{align}
This optimization problem is of a class known as \emph{Bilinear Matrix
Inequality} (BMI) optimization problems \cite{BMIs}.
BMIs are non-convex, and in general, finding a global optimum is an
NP-hard problem \cite{TokerOzbay}.
Nonetheless, for this particular BMI \eqref{origProb}, we are able to
formulate a closed form solution, and to completely specify the optimal
set.

When the dimension $n$ of the quantum system is equal to
the number of possible messages $m$, then perfect retrieval ($P_d=1$)
is achievable by choosing the code-states $\rho_i$ to be mutually
orthogonal pure states, and the measurement such that $\Pi_i=\rho_i$.
When $n<m$ this is no longer possible.
The most straightforward approach to quantum encoding when $n<m$ is
to simply disregard $m-n$ of the messages and aim to perfectly
retrieve the remaining $n$ messages.
It is clear that the smallest probability of error would occur if the
disregarded messages were the ones with smallest prior probabilities.
Thus, this approach is embodied in the ensemble-detector setup
\begin{align}
\Pi_i &= \begin{cases}
            \ket{u_i}\bra{u_i} \quad & 1 \leq i \leq n \\
            0 & n < i \leq m
         \end{cases} \nonumber \\
\rho_i &= \begin{cases}
            \ket{u_i}\bra{u_i} & 1 \leq i \leq n \\
            \text{Don't care} \quad & n < i \leq m
         \end{cases} \label{setup}
\end{align}
where $\{\ket{u_i}\}_{i=1}^n$ is some orthonormal system.
When using this setup $P_d=\sum_{i=1}^n p_i$.

The distinction between classical and quantum systems is very
strongly linked to the fact that non-orthogonality between two
quantum states affects the ability to distinguish between them.
There is no classical analogue of this property.
When the states that a quantum system may be in are mutually
orthogonal, it is said to be in ``the classical limit''.
The fact that the setup \eqref{setup} is comprised only of pure
mutually orthogonal states implies that it is classical in nature
and that the losses encountered are not due to the fact that the
system is governed by quantum mechanics, but to a lossy preprocessing
(disregarding some of the messages).
In the sequel we refer to \eqref{setup} as \emph{pseudo-classical} transmission.

It would, at first glance, seem that one may somehow be able
to utilize the ``quantumness'' of the system, i.e. non-orthogonal
code-states and measurements, in order to improve on the probability
of correct detection $P_d$.
We now formulate and prove a theorem which shows this to be impossible.

\thmbuffer
\begin{theorem}
\label{thm2}
Let $\{p_i\}_{i=1}^m$ be a probability distribution with
$p_1 \geq p_2 \geq \cdots \geq p_m > 0$.
Denoting by $\hat{P}_d$ the maximal probability of correct
detection for a quantum system of dimension $n \leq m$, we
have that
\[
\hat{P}_d = \sum_{i=1}^n p_i.
\]
\end{theorem}
\thmbuffer

\begin{proof}
Let $\tilde{P}_d=\sum_{i=1}^n p_i$.
Since the pseudo-classical setup \eqref{setup} achieves $P_d(\Pi_i,\rho_i)=\tilde{P}_d$,
we have that $\hat{P}_d \geq \tilde{P}_d$.
We prove the theorem by showing that $\hat{P}_d \leq \tilde{P}_d$.

The maximal value of $P_d$ is the solution of \eqref{origProb}.
From Theorem \ref{thm1}, after maximizing with respect to $\rho_i$,
\eqref{origProb} reduces to
\begin{align}
& \max_{\Pi_i} \sum_{i=1}^m p_i \sigma_{\Pi_i}^{\max} \label{EigValProb} \\
& \qquad \st \begin{cases}
            \sumhight\Pi_i \geq 0, & \text{(a)} \\
            \displaystyle\sum_{i=1}^m \Pi_i = I. \qquad & \text{(b)}
                     \end{cases} \nonumber
\end{align}
The constraint (\ref{EigValProb}a) implies that
\begin{equation} \label{relaxation1}
\sigma_{\Pi_i}^{\max} \geq 0,\qquad 1\leq i\leq m,
\end{equation}
and from (\ref{EigValProb}b)
\begin{gather}
\sigma_{\Pi_i}^{\max} \leq 1,\qquad 1\leq i\leq m, \nonumber \\
\sum_{i=1}^m \sigma_{\Pi_i}^{\max} \leq n. \label{relaxation2}
\end{gather}
(The bottom expression in \eqref{relaxation2} is obtained by
taking the trace of (\ref{EigValProb}b)).
We now replace \eqref{EigValProb} by a scalar program,
\begin{align}
& \max_{\sigma_i} \sum_{i=1}^m p_i \sigma_i \label{sigmaProb} \\
& \qquad \st \begin{cases}
            \sumhight 0 \leq \sigma_i \leq 1, \\ 
            \displaystyle\sum_{i=1}^m \sigma_i \leq n. 
                     \end{cases} \nonumber
\end{align}
Problem \eqref{sigmaProb} was created by relaxing the constraints of problem
\eqref{EigValProb} - we keep only the constraints on the eigenvalues and
disregard the original matrix-inequality constraints.
Therefore, the solution of \eqref{sigmaProb} is always larger or equal to
the solution of \eqref{EigValProb}, and thus, serves as an upper bound.

The optimization problem \eqref{sigmaProb} is a linear programme.
Its \emph{Lagrange dual problem} \cite{Boyd&Vanden} is given by
\begin{align}
& \min_{\eta_i,\nu_i,\mu} g(\eta_i,\mu) \label{Dual1} \\
& \qquad \st \begin{cases}
            \eta_i,\nu_i,\mu \geq 0, \quad & \text{(a)} \\
            p_i - \eta_i + \nu_i - \mu = 0, \qquad & \text{(b)}
                     \end{cases} \nonumber
\end{align}
where $ 1\leq i\leq m$ and
\[
g(\eta_i,\mu) = \sum_{i=1}^m \eta_i + n \mu
\]
Using the constraint (\ref{Dual1}b), the variables
$\nu_i$ can be eliminated, yielding
\begin{align}
& \min_{\eta_i,\mu} g(\eta_i,\mu) \label{Dual2} \\
& \qquad \st \begin{cases}
            \eta_i,\mu \geq 0, \\ 
            \eta_i + \mu \geq p_i. 
                     \end{cases} \nonumber
\end{align}

From Lagrange duality theory, for any point in the feasibility
set of \eqref{Dual2}, the objective $g(\eta_i,\mu)$ is greater or
equal to the solution of the primal problem \eqref{sigmaProb}.
In other words, for any dual feasible point $(\eta,\mu)$, $g(\eta_i,\mu)$
is an upper bound on the solution of \eqref{EigValProb}.
Consider
\begin{align}
\hat{\eta}_i &= \begin{cases}
                  p_i - p_{n+1} & 1 \leq i \leq n \\
                  0 & n < i \leq m
               \end{cases} \nonumber \\
\hat{\mu} &= p_{n+1}. \label{guess}
\end{align}
Because $p_1 \geq \cdots \geq p_m$ it is dual feasible.
For this choice
\[
g(\hat{\eta}_i,\hat{\mu}) = \sum_{i=1}^m \hat{\eta}_i + n \hat{\mu}
   = \sum_{i=1}^n (\hat{\eta}_i+\hat{\mu}) = \sum_{i=1}^n p_i.
\]

In conclusion, we have shown that
\[
\max \eqref{origProb} = \max \eqref{EigValProb} \leq \max \eqref{sigmaProb} \leq \min \eqref{Dual2} \leq \sum_{i=1}^n p_i,
\]
which implies that for any valid ensemble and detector
$P_d = \sum_{i=1}^m p_i \Tr( \Pi_i \rho_i ) \leq \tilde{P}_d$.
\end{proof}
\thmbuffer

The implication of Theorem \ref{thm2} is that one can achieve the optimal
probability of correct detection by using 
orthogonal pure states and von Neumann 
measurements, which are easy to implement.
Nevertheless, there may be setups $\{\rho_i,\Pi_i\}_{i=1}^m$ other then
\eqref{setup} which attain $P_d(\Pi_i,\rho_i)=\hat{P}_d$.
In the next section we identify all the ensemble-detector setups which
achieve maximum probability of correct detection.
The importance of characterizing the set of optima is that we may be
able to select an optimum that has preferable performance with regard
to other quality of service measures.
Also, there may be communication protocols which require using a
``non-classical'' ensemble.
These aspects are discussed in greater detail in Section \ref{QKD}.

\section{Characterization of Optimal Setups}
\label{Characterization}

In this section we introduce the notion of \emph{tight frame encoding setups},
and show that all optima are of this form (Theorem \ref{thm3}).
We then fully characterize the set of optima for a given prior probability
distribution (Theorem \ref{thm4} and corollaries).

\subsection{Tight Frame Encoding Setups}

A \emph{tight frame} \cite{YoninaFrames} is a set of $m$ vectors
$\{\ket{u_i}\}_{i=1}^m$ which satisfy
\begin{equation} \label{tightframe}
\sum_{i=1}^m \ket{u_i} \bra{u_i} = I.
\end{equation}
We define a ``Tight Frame Encoding Setup'' (TFES) to be an ensemble-detector
setup of the form
\begin{align}
\Pi_i  &= \ket{u_i}\bra{u_i}, \nonumber \\
\rho_i &= \begin{cases}
            \frac{1}{\bracket{u_i}{u_i}}\ket{u_i}\bra{u_i} \quad & \bracket{u_i}{u_i} > 0 \\
            \text{Don't care}, & \bracket{u_i}{u_i} = 0
          \end{cases} \nonumber
\end{align}
where the vectors $\ket{u_i}$ obey \eqref{tightframe}.
The pseudo-classical setup \eqref{setup} is an example of a TFES.
The probability of correct detection when using a TFES is
$P_d = \sum_{i=1}^m p_i \bracket{u_i}{u_i}$.

The constraint \eqref{tightframe} on the vectors ensures that
$\Pi$ is a valid POVM.
It also implies several properties of the vectors $\ket{u_i}$,
which are summarized in the following lemma.

\thmbuffer
\begin{lemma} \label{TframeLem}
Let $\{\ket{u_i}\}_{i=1}^m$ be a set of vectors which satisfy \eqref{tightframe}.
Then,
\begin{align}
&\sumhight\bracket{u_i}{u_i} \leq 1, \label{normconst} \\
&\text{if } \bracket{u_i}{u_i} = 1 \text{ then } \bracket{u_i}{u_j} = \delta_{i,j}, \label{orthoconst} \\
&\sum_{i=1}^m \bracket{u_i}{u_i} = n. \label{traceconst}
\end{align}
\end{lemma}
\thmbuffer

\begin{proof}
See Appendix \ref{apdx0}.
\end{proof}
\thmbuffer

Tight frames are of interest in many fields and applications where
one seeks a set of vectors whose mutual ``interference'' is minimal.
Specifically, in classical communication, they play an important role in
Syncronous CDMA systems \cite{VerduMultUse,VATcdma}.
Also, the simplex constellation, which is known to be optimal under
certain energy constraints \cite{Balakrishnan,Steiner}, is a tight frame.

The significance of TFESs to quantum encoding is established by the
following result:

\thmbuffer
\begin{theorem} \label{thm3}
All ensemble-detector setups $\{\hat{\rho}_i,\hat{\Pi}_i\}_{i=1}^m$ which
achieve $P_d(\hat{\Pi}_i,\hat{\rho}_i)= \hat{P}_d$ are TFESs.
\end{theorem}
\thmbuffer

The proof of Theorem \ref{thm3}, relies on the following lemma, whose proof
is given in Appendix \ref{apdx1}.

\thmbuffer
\begin{lemma} \label{lem1}
For any ensemble-detector setup $(\hat{\rho}_i,\hat{\Pi}_i)$ which
achieves $P_d(\hat{\Pi}_i,\hat{\rho}_i)=\hat{P}_d$, the largest
eigenvalues of the detection operators satisfy
\[
\sum_{i=1}^m \sigma_{\hat{\Pi}_i}^{\max} = n.
\]
\end{lemma}
\thmbuffer

\begin{proof}(of Thm.\ \ref{thm3})
For any POVM, we have that
\begin{align}
\Tr(\Pi_i) &\geq \sigma_{\Pi_i}^{\max} \label{sumgeqmax} \\
\sum_{i=1}^m \Tr(\Pi_i) &= n \label{lastline}
\end{align}
where \eqref{lastline} comes from taking the trace of the requirement
$\sum_{i=1}^m \hat{\Pi}_i = I$.

Assume that an ensemble-detector setup $\{\hat{\rho}_i,\hat{\Pi}_i\}$ achieves $\hat{P}_d$.
Using \eqref{lastline} with Lemma~\ref{lem1}, we get that
\[
\sum_{i=1}^m \Tr(\hat{\Pi}_i) = \sum_{i=1}^m \sigma_{\hat{\Pi}_i}^{\max},
\]
which, in conjunction with \eqref{sumgeqmax}, shows that for any such detector
\[
\Tr(\hat{\Pi}_i) = \sigma_{\hat{\Pi}_i}^{\max}.\qquad 1 \leq i \leq m
\]
This in turn implies that
\[
\rank(\hat{\Pi}_i) \leq 1,\qquad 1 \leq i \leq m
\]
i.e. the detection elements of any detector which is part of an optimal setup
are of the form $\hat{\Pi}_i=\ket{u_i}\bra{u_i}$ (where $\ket{u_i}$ may also be the
null vector).
In order for $\hat{\Pi}$ to be a valid POVM, the set of vectors $\{\ket{u_i}\}_{i=1}^m$
must obey \eqref{tightframe}.

Since $\{\hat{\rho}_i,\hat{\Pi}_i\}$ is assumed to be an optimal setup, then
$\hat{\rho}_i$ must be an optimal ensemble for the detector $\hat{\Pi}$.
Thus, from Theorem~\ref{thm1} we have that for any $i$, such that $\bracket{u_i}{u_i}>0$,
\[
\hat{\rho}_i=\frac{1}{\bracket{u_i}{u_i}} \ket{u_i}\bra{u_i}.
\]
If $\bracket{u_i}{u_i}=0$, then $\hat{\rho}_i$ can be any quantum state.
\end{proof}
\thmbuffer

An interesting aspect of the above result is that $\hat{P}_d$ can only
be attained by setups in which the detected code-states (those for which
the corresponding measurement operator is not zero) are pure states.
This is hardly surprising, since obviously, for mixed states the chances of
``interference'' between code-states are greater.

\subsection{Choice of TFES}

From Theorem \ref{thm3} we know that all optima are TFESs.
Not all choices of TFES are, however, necessarily optimal.
We now show that the set of optimal TFESs is dependent on the
prior probabilities $\{p_i\}_{i=1}^m$, and on the dimension
of the quantum medium $n$, and characterize this dependance.
The following results (Theorem \ref{thm4} and corollaries) fully
characterize all optimal solutions for a given prior distribution
and dimension $n$.

In order to formulate our results we introduce a classification of
the symbols into three distinct subsets, according to the prior
probability distribution and the dimension $n$.
Recalling that we assume $p_1\geq p_2\geq\cdots\geq p_m$, we define
\begin{itemize}
\item $\mcI_1=\{i\,|\,p_i>p_n\},$
\item $\mcI_2=\{i\,|\,p_i=p_n\},$
\item $\mcI_3=\{i\,|\,p_i<p_n\}.$
\end{itemize}
Note that $\mcI_1$ and $\mcI_3$ may be empty.

\thmbuffer
\begin{theorem} \label{thm4}
Let $\{p_i\}_{i=1}^m$ be a non-increasing distribution of probabilities,
and let $\{\ket{u_i}\}_{i=1}^m$ be the vectors of a TFES in a quantum
system of dimension $n$.
This TFES is optimal in the sense of probability of correct detection
if and only if (i) for all $i \in \mcI_1$, $\bracket{u_i}{u_i}=1$, and (ii)
for all $i \in \mcI_3$, $\bracket{u_i}{u_i}=0$.
\end{theorem}
\thmbuffer

Before proving Theorem \ref{thm4}, we point out the following
important corollaries:

\thmbuffer
\begin{corollary} \label{cor2}
Let $\{p_i\}_{i=1}^m$ be a non-increasing distribution of probabilities, and
let $\{\hat{\rho}_i,\hat{\Pi}_i\}_{i=1}^m$ be an \emph{optimal} encoding
setup in a quantum system of dimension $n$.
Then
\begin{enumerate}
\item $\Pr\{j|i\}=\delta_{i,j} \qquad i \in \mcI_1$,
\item $\Pr\{\det i\}=0 \qquad i \in \mcI_3$.
\end{enumerate}
\end{corollary}
\thmbuffer

\begin{proof}
From Theorem \ref{thm3}, we know that the ensemble-detector setup is a TFES.
From Theorem \ref{thm4}, for all $i \in \mcI_1$, $\hat{\Pi}_i=\ket{u_i}\bra{u_i}$,
such that $\bracket{u_i}{u_i}=1$.
Together with \eqref{orthoconst}, it is easy to see that
\[
\Pr\{j|i\}=\Tr(\hat{\Pi}_j\hat{\rho}_i) = \frac{1}{\bracket{u_i}{u_i}}|\bracket{u_i}{u_j}|^2 = \delta_{i,j}.
\]

Also from Theorem \ref{thm4}, for all $i \in \mcI_3$, $\hat{\Pi}_i=0$, indicating
that the probability of detecting the $i$-th message is
$\Pr\{\det i\}=\sum_j p_j \Tr(\hat{\Pi}_i\hat{\rho}_j)=0$.
\end{proof}

\thmbuffer
\begin{corollary} \label{cor3}
Let $\{p_i\}_{i=1}^m$ be a non-increasing distribution of probabilities.
If $p_n > p_{n+1}$ then any optimal setup $\{\hat{\rho}_i,\hat{\Pi}_i\}_{i=1}^m$
must be of the form \eqref{setup} (pseudo-classical).
\end{corollary}
\thmbuffer

\begin{proof}
From Theorem \ref{thm3}, the optimal setup must be a TFES.
When $p_n>p_{n+1}$ we have $\mcI_3=\{n+1,\ldots,m\}$, which,
using Theorem \ref{thm4}, indicates that
\[
\bracket{u_i}{u_i} = 0 \qquad n+1 \leq i \leq m.
\]
Together with \eqref{tightframe} this implies
\begin{equation} \label{orthonormal}
\sum_{i=1}^n \ket{u_i}\bra{u_i} = I.
\end{equation}
A set of $n$ vectors in $n$-dimensional space can satisfy \eqref{orthonormal} if and only
if they form an orthonormal set.
Thus, the only optimal setup when $p_n>p_{n+1}$ is \eqref{setup}.
\end{proof}

\thmbuffer
\begin{corollary} \label{cor4}
If $p_i = \frac{1}{m}$ for all $i$, then all TFESs achieve $\hat{P}_d$.
\end{corollary}
\thmbuffer

\begin{proof}
The Corollary follows directly from Theorem \ref{thm4}, for $\mcI_1=\mcI_3=\emptyset$.
\end{proof}
\thmbuffer

We now prove Theorem \ref{thm4}.

\thmbuffer
\begin{proof}(of Thm.\ \ref{thm4})
Assume that $\{\ket{u_i}\}_{i=1}^m$ are the vectors of a TFES
which is optimal in the sense of $P_d$.

Assume that $\mcI_1\neq\emptyset$ and denote by $k$ 
the largest index in $\mcI_1$.
(i.e. $\mcI_1=\{1,\ldots,k\}$).
This means that $p_k>p_{k+1}=p_{k+2}=\cdots=p_n$ (from the
definition of $\mcI_1$, we have that $k < n$).
For any TFES we can write
\begin{align}
P_d &= \sum_{i=1}^m p_i \bracket{u_i}{u_i} \nonumber \\
    &= \sum_{i=1}^k p_i \bracket{u_i}{u_i} + \sum_{i=k+1}^m p_i \bracket{u_i}{u_i} \nonumber \\
    &\leq \sum_{i=1}^k p_i \bracket{u_i}{u_i} + p_{k+1} \sum_{i=k+1}^m \bracket{u_i}{u_i} \label{Pd1b} \\
    &= \sum_{i=1}^k p_i \bracket{u_i}{u_i} + p_{k+1} \left( n - \sum_{i=1}^k \bracket{u_i}{u_i} \right) \label{Pd2b} \\
    &= \sum_{i=1}^k \big[ p_i \bracket{u_i}{u_i} + p_{k+1} (1-\bracket{u_i}{u_i}) \big] + (n-k) p_{k+1}, \label{Pd3b}
\end{align}
where the transition from \eqref{Pd1b} to \eqref{Pd2b} relies on \eqref{traceconst}.

Recall that for all $i \in \mcI_1$ we have $p_i>p_{k+1}$.
If for some $1 \leq i \leq k$, $\bracket{u_i}{u_i} < 1$, then from \eqref{Pd3b}
\begin{align}
P_d &< \sum_{i=1}^k \big[ p_i \bracket{u_i}{u_i} + p_i (1 - \bracket{u_i}{u_i}) \big] + (n-k) p_{k+1} \nonumber \\
    &= \sum_{i=1}^k p_i + (n-k) p_{k+1} = \sum_{i=1}^n p_i = \hat{P}_d. \nonumber
\end{align}
Here we have relied on the fact that $p_{k+1}=p_{k+2}=\cdots=p_n$.
Therefore, in order to achieve $\hat{P}_d$, the vectors $\ket{u_i}$ must satisfy
\[
\bracket{u_i}{u_i} = 1, \qquad i \in \mcI_1
\]
This concludes the proof of the first statement of the `only if' direction.

We go on to prove the second statement.
Assume that $k'~=~\max\mcI_2<m$ (i.e. $\mcI_3=\{k'+1,\ldots,m\}\neq\emptyset$).
By definition $p_{k'}~>~p_{k'+1}$. 
We again have
\[
P_d = \sum_{i=1}^m p_i \bracket{u_i}{u_i}
    = \sum_{i=1}^{k'} p_i \bracket{u_i}{u_i} + \sum_{i=k'+1}^m p_i \bracket{u_i}{u_i}.
\]
If for some $i \in \mcI_3$, $\bracket{u_i}{u_i}>0$, then
\begin{align}
P_d &< \sum_{i=1}^{k'} p_i \bracket{u_i}{u_i} + p_{k'} \sum_{i=k'+1}^m \bracket{u_i}{u_i} \label{Pd0a} \\
    &= \sum_{i=1}^n p_i \bracket{u_i}{u_i} + p_n \sum_{i=n+1}^m \bracket{u_i}{u_i} \label{Pd1a} \\
    &= \sum_{i=1}^n p_i \bracket{u_i}{u_i} + p_n \left( n - \sum_{i=1}^n \bracket{u_i}{u_i} \right) \label{Pd2a} \\
    &= \sum_{i=1}^n p_i \bracket{u_i}{u_i} + p_n \sum_{i=1}^n \big( 1 - \bracket{u_i}{u_i} \big) \nonumber\\
    &\leq \sum_{i=1}^n p_i \bracket{u_i}{u_i} + \sum_{i=1}^n p_i \big( 1 - \bracket{u_i}{u_i} \big) \nonumber \\
    &= \sum_{i=1}^n p_i = \hat{P}_d, \nonumber
\end{align}
where the transitions from \eqref{Pd0a} to \eqref{Pd2a} rely on the fact that
$k' \in \mcI_2$ and on \eqref{traceconst}.
Thus, for any TFES which achieves maximal $P_d$
\[
\bracket{u_i}{u_i} = 0, \qquad i \in \mcI_3.
\]

We continue by proving the `if' direction.
Assume that for all $i \in \mcI_1$, $\bracket{u_i}{u_i}=1$, and that
for all $i \in \mcI_3$, $\bracket{u_i}{u_i}=0$.
We must first note that under these conditions, using
\eqref{traceconst} yields
\begin{equation} \label{traceconst2}
\sum_{i=1}^{k'} \bracket{u_i}{u_i} = n.
\end{equation}
If $\mcI_1=\emptyset$, then $\mcI_2=\{1,\dots,k'\}$.
We can then write
\begin{align}
P_d &= \sum_{i=1}^m p_i \bracket{u_i}{u_i} \label{Pd1c} \\
    &= p_1 \sum_{i=1}^{k'} \bracket{u_i}{u_i} \label{Pd2c} \\
    &= n p_1 = \sum_{i=1}^n p_i = \hat{P}_d, \label{Pd3c}
\end{align}
where the transition from \eqref{Pd1c} to \eqref{Pd2c} relies
on the facts that for all $i>k'$, $\bracket{u_i}{u_i}=0$, and $p_1=p_2=\cdots=p_n$.
The transition from \eqref{Pd2c} to \eqref{Pd3c} is based on \eqref{traceconst2}.

If $\mcI_1=\{1,\dots,k\}$ and $\mcI_2=\{k+1,\dots,k'\}$ then, similarly,
\begin{align}
P_d &= \sum_{i=1}^m p_i \bracket{u_i}{u_i} \nonumber \\
    &= \sum_{i=1}^{k} p_i \bracket{u_i}{u_i} + p_{k+1} \sum_{i=k+1}^{k'} \bracket{u_i}{u_i} \nonumber \\
    &= \sum_{i=1}^{k} p_i + (n-k) p_{k+1} = \sum_{i=1}^n p_i = \hat{P}_d, \nonumber
\end{align}
thereby completing the proof.
\end{proof}
\thmbuffer

Theorem \ref{thmBig} below summarizes the assertions of Theorems
\ref{thm2}, \ref{thm3} and \ref{thm4}, in concise form, and completely
characterizes all optimal transmitter-receiver setups.

\thmbuffer
\begin{theorem} \label{thmBig}
Let $\{p_i\}_{i=1}^m$ be a probability distribution with
$p_1 \geq p_2 \geq \cdots \geq p_m > 0$.
For a given number $n \leq m$, define the index sets
\begin{itemize}
\item $\mcI_1=\{i\,|\,p_i>p_n\},$
\item $\mcI_2=\{i\,|\,p_i=p_n\},$
\item $\mcI_3=\{i\,|\,p_i<p_n\}.$
\end{itemize}
The maximal probability of correct detection for a quantum
system of dimension $n \leq m$ is
\[
\hat{P}_d = \sum_{i=1}^n p_i.
\]
The optimum is achieved if and only if the ensemble-detector
setup is of the form
\begin{align}
\Pi_i  &= \ket{u_i}\bra{u_i}, \nonumber \\
\rho_i &= \begin{cases}
            \frac{1}{\bracket{u_i}{u_i}}\ket{u_i}\bra{u_i} \quad & \bracket{u_i}{u_i} > 0 \\
            \text{Don't care}, & \bracket{u_i}{u_i} = 0
          \end{cases} \nonumber
\end{align}
where the vectors $\{\ket{u_i}\}_{i=1}^m$ obey
\begin{align}
\sum_{i=1}^m \ket{u_i}\bra{u_i} &= I, \nonumber \\
\bracket{u_i}{u_i} &= 1, \qquad i\in\mcI_1 \nonumber \\
\bracket{u_i}{u_i} &= 0. \qquad i\in\mcI_3 \nonumber
\end{align}
\end{theorem}

Put in words, maximum $P_d$ can only be attained by a TFES, where
the messages with high prior probabilities ($i\in\mcI_1$) are encoded
using orthogonal code states, and are thus recovered perfectly
(Corollary~\ref{cor2}), and the messages with low prior probabilities
($i\in\mcI_3$) are discarded - much like in pseudo-classical encoding.
In choosing the remaining frame vectors, one has freedom and they can be
chosen to be non-orthogonal.
Important special cases are when $p_{n+1}>p_n$, where one has no freedom
and the only optimum is pseudo-classical encoding (Corollary~\ref{cor3}),
and the equiprobable case $p_i = \frac{1}{m}$, where there is complete
freedom in choosing the TFES frame vectors (Corollary~\ref{cor4}).

\subsection{Application to the Analysis of Communication Protocols}
\label{QKD}

In many applications, additional constraints, other then the ones imposed
by the physics, are placed on the encoding-retrieval setup.
In quantum key distribution \cite{QKDtutorial}, for example, constraints
arise due to the need for security against eavesdropping.
Further constraints may occur due to technical (implementation) issues.
The work at hand can then serve for two purposes.
The first is to quantify the degradation in $P_d$ due to the need to meet
the extra design constraints.
This can be done by simply comparing the performance of the constrained
system to the theoretical upper bound $\hat{P}_d$.
The second possible use of this work, in this context, is to search within
the set of optimal TFESs for a setup, which is close to meeting the
demands posed by the application.
When taking the latter approach we are assured optimal performance with
regard to $P_d$.

Consider the BB84 protocol \cite{BB84}.
In this QKD protocol, Alice wishes to send Bob secure binary information.
In order to counter possible eavesdropping, she sends one of $m=4$ messages
with $p_i=\frac{1}{4}$ over a 2-dimensional quantum channel.
The code-states used are denoted $\ket{u_{ij}}$, where $i,j=0,1$, and they
obey the relations
\begin{align}
|\bracket{u_{ij}}{u_{i'j'}}|^2 &= \begin{cases}
                                \delta_{j,j'} & i=i' \\
                                1/2 & i \neq i' \end{cases} \nonumber \\
\frac{1}{2} \sum_{i,j=0}^1 \ket{u_{ij}}\bra{u_{ij}} &= I \label{BB84frame}
\end{align}
Note that \eqref{BB84frame} indicates that this collection of vectors is a tight frame.

Bob utilizes the POVM (of order 4) $\Pi_{ij}=\frac{1}{2}\ket{u_{ij}}\bra{u_{ij}}$,
in order to retrieve Alice's message.
They then exchange knowledge on which ``pair of states'' was received (by, for
example, comparing the $i$ index).
If both the sent and the detected symbols originate from the same pair, then
the transferred bit of information is taken as the member of the pair that was
detected (the $j$ index).
If the symbols originate from different pairs, the received symbol is discarded.
In order to promote security, Alice and Bob use $m>n$, at a cost of reduced
data rate.
The security of this protocol has been extensively studied.

The probability of correct detection achieved by Bob prior to the exchange
of the $i$ index is $P_d=1/2$.
This is equal to the upper bound $\hat{P}_d$ for this case, meaning that
under the requirement of countering eavesdropping, Bob achieves the maximal
possible performance.
The fact that the upper bound is reached would hardly surprise most readers,
in the context of a protocol as simple as BB84.
It does, however, serve to illustrate the possible use of the unconstrained
upper bound $\hat{P}_d$ in quantifying the efficacy of more complex
communication protocols.

\section{Optimal Worst-Case Posterior Probability}
\label{posterior}

An alternative quality of service measure for systems of digital
communication/storage is the \emph{worst-case posterior probability}
\cite{CoxHi,KWER}.
The posterior probability, defined as
\begin{align}
P_p(i) &\triangleq \frac{\Pr\{\text{message $i$ detected correctly}\}}{\Pr\{\text{message $i$ detected}\}} \nonumber \\
    &= \frac{p_i \Tr(\Pi_i \rho_i)}{\sum_j p_j \Tr(\Pi_i \rho_j)}, \label{posterDef}
\end{align}
is the answer to the question: ``Given that the detected message is $i$,
what is the probability that it is the right answer?''.
The worst-case posterior probability is then
\[
P_p \triangleq \min_{i=1,\ldots,m} P_p(i).
\]
The higher the value of $P_p$, the more reliable the output of the
measurement.

Denote $\Pr\{\det i\}=\sum_j p_j \Tr(\Pi_i \rho_j)$ the probability
of detecting the $i$-th outcome.
By definition
\begin{equation} \label{ineq2000}
P_p \cdot \Pr\{\text{det }i\} \leq p_i \Pr\{i|i\}. \qquad 1 \leq i \leq m
\end{equation}
Summing the inequalities \eqref{ineq2000} over $i$, one gets
\[
P_p = P_p \cdot \sum_{i=1}^m \Pr\{\text{det }i\} \leq P_d.
\]
Thus, in any digital encoding system (not necessarily quantum mechanical) the
value of $P_p$ is bounded above by the value of $P_d$.
In particular, for quantum systems, this means that a universal upper bound on
$P_p$ is $\hat{P}_d$.
Theorem \ref{thm5} below provides a simple method for finding an upper
bound on $P_p$ for a given set of code-states $\rho_i$ and prior
probabilities $p_i$.
We present an example in which our bound is tighter than the universal
bound $\hat{P}_d$.

Obtaining the optimal measurement in the sense of $P_p$ involves
a bisection procedure, where each step is computationally expensive
(solving an SDP) \cite{KWER}.
The bound obtained using our method can serve to shorten the initial
bisection interval, thereby reducing the computational cost of finding
the optimal detector.
We also hope that our method can serve to find tighter universal upper
bounds on $P_p$.

Note that for the pseudo-classical TFES \eqref{setup}, and in fact
for any setup in which one of the POVM elements is zero, the posterior
probability is ill-defined, since the denominator in \eqref{posterDef}
is zero.
We therefore introduce a surrogate measure of the reliability of the
outcome, designed to replace $P_p$ in this case.

Since we seek a measure of reliability of the output, there is no
point in taking into account outputs which never occur.
Hence we choose to measure the most unreliable outcome, of the
set of \emph{possible} outcomes.
The \emph{effective worst-case posterior probability} is defined as
\[
P_p^{\eff} = \min_{i|\Pr\{\det i\}>0} P_p(i).
\]
Note that whenever $P_p$ is well defined $P_p^{\eff} = P_p$.
In addition, the upper bound $P_d$ also holds for $P_p^{\eff}$.

According to Theorem~\ref{thmBig}, in many cases, ensemble-detector setups
which attain optimal $P_d$ are a TFES, whose detector has zero elements.
For all $i$, such that $\Pi_i=0$, we can choose the code states freely,
without degrading the performance in $P_d$.
This raises the question, how should one choose the `don't care' states,
so that the output of the system would be reliable?
We show that for the pseudo-classical TFES \eqref{setup}, there is a
choice of `don't care' states which attains the maximum value of $P_p^{\eff}$.

\subsection{An Upper Bound on $P_p$ for a Given Ensemble}

\thmbuffer
\begin{theorem} \label{thm5}
Let $\{\rho_i\}_{i=1}^m$ be $m$ arbitrary quantum states of
dimension $n$, with prior probabilities $p_i$.
Define the operators
\[
A_i(\delta) = (1-\delta) \sum_{k=1}^m p_k \rho_k - p_i \rho_i,
\]
where $\delta\in\mcR$.
If for some $1 \leq i \leq m$, $A_i(\delta) \geq 0$, then
$P_p \leq 1-\delta$.
\end{theorem}
\thmbuffer

\begin{proof}
Assume that $\{\rho_i\}_{i=1}^m$ is an arbitrary ensemble
of quantum states with prior probabilities $p_i$.
Denote by $\hat{\Pi}$ and $\hat{s}$ the solution to
\begin{align}
& \min_{\Pi_i,s} s \label{feasProb} \\
& \qquad \st \begin{cases}
            \Pi_i \geq 0, \\
            \displaystyle\sum_{i=1}^m \Pi_i = I \\
            \Tr [\Pi_i A_i (\delta)] \leq s, \quad 1 \leq i \leq m
                     \end{cases} \nonumber
\end{align}
In \cite{KWER} it was shown that if the value of \eqref{feasProb}
is non-negative (i.e. $\hat{s}\geq0$) for a specific choice of
$\delta$, then\footnote{Actually, the authors of \cite{KWER} are
concerned with an error function which is equal to $1-P_p$.}
$P_p(\hat{\Pi}) \leq 1-\delta$.
This statement contains a slight inaccuracy, because for $P_p(\hat{\Pi})$
to be well-defined, one must also include in \eqref{feasProb} the
constraint $\Pi_i\neq0$ (the authors do mention `taking a short cut').

The dual program of \eqref{feasProb} is
\begin{align}
& \max_{Y,\lambda_i} \Tr(Y) \nonumber \\
& \qquad \st \begin{cases}
            \lambda_i \geq 0, \\
            \displaystyle\sum_{i=1}^m \lambda_i = 1, \\
            \lambda_i A_i(\delta) - Y \geq 0.
                     \end{cases} \nonumber
\end{align}
This means that if, for a specific value of $\delta$, one can
find real scalars $\lambda_i$ and an operator $Y$, such that
\begin{align}
\lambda_i & \geq 0, \nonumber \\
\sum_{i=1}^m \lambda_i &= 1, \nonumber \\
\lambda_i A_i(\delta) - Y &\geq 0, \label{dualReq}
\end{align}
then $\Tr(Y) \leq \hat{s}$.
Therefore, if in addition to the requirements \eqref{dualReq},
$Y$ also satisfies $\Tr(Y) \geq 0$, then we are assured that
$\hat{s} \geq 0$, and that $1-\delta$ is an upper bound on the
\emph{optimal} posterior probability $\hat{P}_p$.

Define the index subset
\[
Q(\delta) = \left\{ i | A_i(\delta) \geq 0 \right\},
\]
and denote its cardinality by $|Q(\delta)|$.
If $Q(\delta)$ is non-empty, then we can choose
\[
Y = 0, \qquad \quad \lambda_i = \begin{cases}
                    0, & i \notin Q(\delta) \\
                    \frac{1}{|Q(\delta)|}, & i \in Q(\delta) \end{cases}
\]
which satisfy all the above requirements \eqref{dualReq}.
Thus, whenever $Q(\delta)$ is non-empty, $1-\delta$ is an upper bound on
$\hat{P}_p$.
\end{proof}
\thmbuffer

As an example of the application of Theorem~\ref{thm5}, we examine an ensemble
comprised of the pure states $\rho_i^0 = \ket{u_i} \bra{u_i}$ in a two dimensional
Hilbert space,
\[ 
u_1 = \begin{pmatrix} 0 \\  1  \end{pmatrix} \qquad
u_2 = \frac{1}{2} \begin{pmatrix} \sqrt{3} \\ -1 \end{pmatrix} \qquad
u_3 = \frac{1}{2} \begin{pmatrix} \sqrt{3} \\  1 \end{pmatrix}
\] 
with prior probabilities
\[ 
p_1 = 0.4 \qquad\qquad p_2 = p_3 = 0.3.
\] 
For this ensemble,
\[
A_2(\delta) = \frac{1}{40} \begin{pmatrix} 9-18\delta & \sqrt{27} \\
                                           \sqrt{27}  & 19-22\delta \end{pmatrix}
\]
whose eigenvalues are
\[
\sigma_{A_2} = \frac{1}{20} \big( 7 - 10\delta \pm \sqrt{13-5\delta+\delta^2} \big).
\]
This implies that $A_2(\delta)$ is PSD for any $\delta\leq0.36$, and thus the upper
bound provided by Theorem~\ref{thm5} is $P_p \leq 0.64$.
This is an improvement over the universal bound $\hat{P}_d=0.7$.

\subsection{Choosing the `Don't Care' States of Optimal TFESs}

In many situations, setups which attain maximum $P_d$, have $\Pi_i=0$
for some $i$.
When this is the case, there are undecided degrees of freedom to the
TFES - the `don't care' states.
We would like to be able to choose these states so that the outcome
of the measurement is reliable.
We measure the reliability using $P_p^{\eff}$ defined above.

We present a choice of `don't care' states for the pseudo-classical setup
for which $P_p^{\eff} = P_d$, i.e. when the pseudo-classical setup is used
with this choice of `don't care' states, its performance is optimal both in
terms of $P_d$ and in terms of $P_p^{\eff}$.

\thmbuffer
\begin{theorem} \label{cor6}
When using the pseudo-classical TFES \eqref{setup}, with the choice
\[
\rho_j = \frac{1}{\sum_{i=1}^n p_i} \sum_{i=1}^n p_i\ket{u_i}\bra{u_i},\qquad j=n+1,\ldots,m
\]
for the `don't care' states, $P_p^{\eff}$ attains the upper bound $P_d$.
\end{theorem}
\thmbuffer

\begin{proof}
For all $i\leq n$ we get
\begin{align}
P_p(i) &= \frac{p_i}{p_i + \sum_{j=n+1}^m \frac{p_j}{\sum_{i=1}^n p_i} \sum_{k=1}^n p_k|\bracket{u_i}{u_k}|^2} \nonumber \\
       &= \frac{p_i}{p_i + \sum_{j=n+1}^m \frac{p_i p_j}{\sum_{i=1}^n p_i}} \nonumber \\
       &= \frac{\sum_{i=1}^n p_i}{\sum_{i=1}^n p_i+\sum_{j=n+1}^m p_j} \nonumber \\
       &= \sum_{i=1}^n p_i. \nonumber
\end{align}
Thus $P_p^{\eff} = \sum_{i=1}^n p_i = \hat{P}_d$.
\end{proof}

\section{Conclusion}
\label{conclusion}
We have addressed the question of retrieval of digital data encoded
in a quantum medium, using as our main performance criterion the
probability of correct detection.
We have found the optimal code-states for an arbitrary detector,
and the optimal encoding-retrieval setups for an arbitrary prior
distribution.

In terms of $P_d$ one cannot do better then pseudo-classical
transmission (orthonormal code-states and measurement operators).
We have also shown that of all the setups which attain maximal $P_d$,
the pseudo-classical TFES can be made to have optimal effective worst-case
posterior probability.
We have, however, indicated that under certain circumstances, there are
benefits for using fully quantum setups (non-orthogonal code-states).

The natural extension of this work is the design of optimal setups with
added constraints.
Such constraints may arise due to requirements other than reliable
communication, such as the need for security discussed above.
Constraints may also stem from implementation issues which are typical
to specific quantum systems that regularly serve for transmission and
storage of information.

\section*{Acknowledgements}

We would like to thank Moshe Nazarathy, who helped spark our interest in the
subjects of this paper.
We also benefited from discussions with Oded Regev and Tal Mor.

\useRomanappendicesfalse
\renewcommand{\theequation}{\thesection.\arabic{equation}}

\appendices
\setcounter{equation}{0}
\section{Proof of Lemma \ref{TframeLem}} \label{apdx0}
For all $1\leq i\leq m$ we have that
\[
|\bracket{u_i}{u_i}|^2 \leq \sum_{k=1}^m |\bracket{u_i}{u_k}|^2 = \bra{u_i} \sum_{k=1}^m \ket{u_k}\bracket{u_k}{u_i}.
\]
Using \eqref{tightframe}, this implies that
\[
|\bracket{u_i}{u_i}|^2 \leq \bra{u_i}I\ket{u_i} = \bracket{u_i}{u_i}.
\]
Thereupon $\bracket{u_i}{u_i} \leq 1$, proving the property \eqref{normconst}.

If $\bracket{u_i}{u_i} = 1$ then
\[
\bracket{u_i}{u_i} = \bra{u_i} \sum_{k=1}^m \ket{u_k}\bracket{u_k}{u_i} = \sum_{k=1}^m |\bracket{u_i}{u_k}|^2 = 1
\]
making
\[
\sum_{k \neq i} |\bracket{u_i}{u_k}|^2 = 0
\]
Since this is a sum of nonnegative numbers, then for all $k \neq i$ we have
\[
|\bracket{u_i}{u_k}|^2 = 0 \qquad\Rightarrow\qquad \bracket{u_i}{u_k} = 0
\]
proving \eqref{orthoconst}.
Property \eqref{traceconst} follows from taking the trace of \eqref{tightframe}.

\setcounter{equation}{0}
\section{Proof of Lemma \ref{lem1}} \label{apdx1}
Assume that $(\bar{\eta}_i,\bar{\mu})$ is a feasible point of the
programme \eqref{Dual2}, such that $\bar{\mu}=0$.
From the constraint (\ref{Dual2}b), $\bar{\eta}_i$ must satisfy
$\bar{\eta}_i \geq p_i$ and then
\[
g(\bar{\eta}_i,\bar{\mu}) \geq \sum_{i=1}^m p_i > \sum_{i=1}^n p_i = g(\hat{\eta}_i,\hat{\mu}),
\]
where $(\hat{\eta}_i,\hat{\mu})$ are defined in \eqref{guess}.
Thus $(\bar{\eta}_i,\bar{\mu})$ cannot be a dual optimal point.
All dual optimal points must satisfy $\mu \neq 0$.

One of the KKT conditions for the solution to problem \eqref{sigmaProb} is
\[
\mu \left( n - \sum_{i=1}^m \sigma_i \right) = 0.
\]
Since the dual optimal $\mu \neq 0$, then any optimal values of
$\sigma_i$ must satisfy.
\begin{equation} \label{kkt2}
\sum_{i=1}^m \sigma_i = n.
\end{equation}

Let $\{\Pi_i\}$ be a POVM, which is part of an optimal ensemble-detector setup, i.e.
$\sum_i p_i \sigma_{\Pi_i}^{\max} = \hat{P}_d$.
By choosing
\begin{equation} \label{sigDefine}
\hat{\sigma}_i = \sigma_{\Pi_i}^{\max}
\end{equation}
we get $\sum_i p_i \hat{\sigma}_i = \hat{P}_d$, ensuring that $\hat{\sigma}_i$ are an
optimum of \eqref{sigmaProb}, and thus satisfy \eqref{kkt2}.
In conjunction with \eqref{sigDefine}, this proves the Lemma.


\bibliographystyle{IEEEtran}
\bibliography{EncodingBibliography}

\begin{thebibliography}{10}
\providecommand{\url}[1]{#1}
\csname url@rmstyle\endcsname
\providecommand{\newblock}{\relax}
\providecommand{\bibinfo}[2]{#2}
\providecommand\BIBentrySTDinterwordspacing{\spaceskip=0pt\relax}
\providecommand\BIBentryALTinterwordstretchfactor{4}
\providecommand\BIBentryALTinterwordspacing{\spaceskip=\fontdimen2\font plus
\BIBentryALTinterwordstretchfactor\fontdimen3\font minus
  \fontdimen4\font\relax}
\providecommand\BIBforeignlanguage[2]{{%
\expandafter\ifx\csname l@#1\endcsname\relax
\typeout{** WARNING: IEEEtran.bst: No hyphenation pattern has been}%
\typeout{** loaded for the language `#1'. Using the pattern for}%
\typeout{** the default language instead.}%
\else
\language=\csname l@#1\endcsname
\fi
#2}}

\bibitem{Peres:BOOK}
A.~Peres, \emph{Quantum Theory: Concepts and Methods}.\hskip 1em plus 0.5em
  minus 0.4em\relax Waterloo, Canada: Kluwer Academic Publishers, 1993.

\bibitem{CoverThomas}
T.~M. Cover and J.~A. Thomas, \emph{Elements of Information Theory}.\hskip 1em
  plus 0.5em minus 0.4em\relax John Wiley \& sons, 1991.

\bibitem{QKDtutorial}
N.~Gisin, G.~Ribordy, W.~Tittel, and H.~Zbinden, ``Quantum cryptography,''
  \emph{Rev. Mod. Phys.}, vol.~74, pp. 145--195, Jan 2002.

\bibitem{Holevo1}
A.~S. Holevo, ``Statistical decision theory for quantum systems,'' \emph{J.
  Multivar. Anal.}, vol.~3, pp. 337--394, Dec 1973.

\bibitem{YKL1}
H.~P. Yuen, R.~S. Kennedy, and M.~Lax, ``Optimum testing of multiple hypotheses
  in quantum detection theory,'' \emph{IEEE Trans. Inform. Theory}, vol. IT-21,
  pp. 125--134, 1975.

\bibitem{YoninaMegretskiVerghese1}
Y.~C. Eldar, A.~Megretski, and G.~C. Verghese, ``Designing optimal quantum
  detectors via semidefinite programming,'' \emph{IEEE Trans. Inform. Theory},
  vol.~49, pp. 1012--1017, 2003.

\bibitem{Helstrom1}
C.~W. Helstrom, \emph{Quantum Detection and Estimation Theory}.\hskip 1em plus
  0.5em minus 0.4em\relax New York: Academic Press, 1976.

\bibitem{CBH}
M.~Charbit, C.~Bendjaballah, and C.~W. Helstrom, ``Cutoff rate for the $m$-ary
  psk modulation channel with optimal quantum detection,'' \emph{IEEE Trans.
  Inform. Theory}, vol.~35, pp. 1131--1133, Sep 1989.

\bibitem{OBH}
M.~Osaki, M.~Ban, and O.~Hirota, ``Derivation and physical interpretation of
  the optimum detection operators for coherent-state signals,'' \emph{Phys.
  Rev. A}, vol.~54, pp. 1691--1701, Aug 1996.

\bibitem{BKMH}
M.~Ban, K.~Kurokawa, R.~Momose, and O.~Hirota, ``Optimum measurements for
  discrimination among symmetric quantum states and parameter estimation,''
  \emph{Int. J. Theor. Phys.}, vol.~36, pp. 1269--1288, 1997.

\bibitem{YoninaForney1}
Y.~C. Eldar and G.~D. {Forney, Jr.}, ``On quantum detection and the square-root
  measurement,'' \emph{IEEE Trans. Inform. Theory}, vol.~47, pp. 858--872,
  2001.

\bibitem{YoninaMegretskiVerghese2}
Y.~C. Eldar, A.~Megretski, and G.~C. Verghese, ``Optimal detection of symmetric
  mixed quantum states,'' \emph{IEEE Trans. Inform. Theory}, vol.~50, pp.
  1198--1207, 2004.

\bibitem{KWER}
\BIBentryALTinterwordspacing
R.~L. Kosut, I.~Walmsley, Y.~C. Eldar, and H.~Rabitz, ``Quantum state detector
  design: Optimal worst-case a posteriori performance,'' 2004, submitted for
  publication. [Online]. Available: \url{http://arXiv.org/abs/quant-ph/0403150}
\BIBentrySTDinterwordspacing

\bibitem{Ivanovic}
I.~D. Ivanovic, ``How to differentiate between non-orthogonal states,''
  \emph{Phys. Lett. A}, vol. 123, pp. 257--259, Aug 1987.

\bibitem{YoninaUnambig}
Y.~C. Eldar, ``A semidefinite programming approach to optimal unambiguous
  discrimination of quantum states,'' \emph{IEEE Trans. Inform. Theory},
  vol.~49, pp. 446--456, Feb 2003.

\bibitem{YoninaStojnicHassibi}
Y.~C. Eldar, M.~Stojnic, and B.~Hassibi, ``Optimal quantum detectors for
  unambiguous detection of mixed states,'' \emph{Phys. Rev. A}, vol.~69, no.~6,
  p. 062318, 2004.

\bibitem{SaMoHi}
M.~Sasaki, R.~Momose, and O.~Hirota, ``Quantum detection for on-off keyed
  mixed-state signals with a small amount of thermal noise,'' \emph{Phys. Rev.
  A.}, vol.~55, no.~4, pp. 3222--3225, Apr 1997.

\bibitem{VilnrotterLau2}
\BIBentryALTinterwordspacing
V.~Vilnrotter and C.~W. Lau, ``Quantum detection of binary and ternary signals
  in the presence of thermal noise fields,'' \emph{The InterPlanetary Network
  Progress Report 42-152, October-Decmber 2002}, Feb 2003. [Online]. Available:
  \url{http://ipnpr.jpl.nasa.gov/tmo/progress\_report/42-152/152B.pdf}
\BIBentrySTDinterwordspacing

\bibitem{ConchaPoor}
J.~I. Concha and H.~V. Poor, ``Multiaccess quantum channels,'' \emph{IEEE
  Trans. Inform. Theory}, vol.~50, pp. 725--747, May 2004.

\bibitem{NoamYonina1}
\BIBentryALTinterwordspacing
N.~Elron and Y.~C. Eldar, ``Quantum detection with uncertain states,''
  \emph{Phys. Rev. A.}, vol.~72, p. 032338, 2005. [Online]. Available:
  \url{http://arxiv.org/abs/quant-ph/0501084}
\BIBentrySTDinterwordspacing

\bibitem{DaSK}
\BIBentryALTinterwordspacing
G.~M. {D'Ariano}, M.~F. Sacchi, and J.~Kahn, ``Minimax quantum state
  discrimination,'' 2005, submitted to Phys. Rev. A. [Online]. Available:
  \url{http://arXiv.org/abs/quant-ph/0504048}
\BIBentrySTDinterwordspacing

\bibitem{BMIs}
\BIBentryALTinterwordspacing
Y.~Xiao, F.~Crusca, and E.~{K.-w. Chu}, ``Bilinear matrix inequalities in
  robust control: Phase {I} - problem formulation,'' Monash University,
  Victoria, Australia, Tech. Rep. MECSE-3-1996, Apr 1996. [Online]. Available:
  \url{http://www.ds.eng.monash.edu.au/techrep/reports/}
\BIBentrySTDinterwordspacing

\bibitem{TokerOzbay}
O.~Toker and H.~{\"{O}zbay}, ``On the {$\mathcal{NP}$}-hardness of solving
  bilinear matrix inequalities and simultaneous stabilization with static
  output feedback,'' in \emph{Proc. of the 1995 American Control Conference},
  Seattle, Washington, 1995, pp. 2525--2526.

\bibitem{Boyd&Vanden}
S.~Boyd and L.~Vandenberghe, \emph{Convex Optimization}.\hskip 1em plus 0.5em
  minus 0.4em\relax Cambridge University Press, Mar 2004.

\bibitem{YoninaFrames}
Y.~C. Eldar and G.~D. {Forney, Jr.}, ``Optimal tight frames and quantum
  measurement,'' \emph{IEEE Trans. Inform. Theory}, vol.~48, no.~3, pp.
  599--610, Mar 2002.

\bibitem{VerduMultUse}
S.~{Verd\'{u}}, \emph{Multiuser Detection}.\hskip 1em plus 0.5em minus
  0.4em\relax Cambridge University Press, 1998.

\bibitem{VATcdma}
P.~Viswanath, V.~Anantharam, and D.~N.~C. Tse, ``Optimal sequences, power
  control, and user capacity of syncronous {CDMA} systems with linear {MMSE}
  multiuser receivers,'' \emph{IEEE Trans. Inform. Theory}, vol.~45, no.~6, pp.
  1968--1983, Sep 1999.

\bibitem{Balakrishnan}
A.~V. Balakrishnan, ``A contribution to the sphere-packing problem of
  communication systems,'' \emph{J. Math. Anal. Appl.}, vol.~3, pp. 485--506,
  Dec 1961.

\bibitem{Steiner}
M.~Steiner, ``The strong simplex conjecture is false,'' \emph{IEEE Trans.
  Inform. Theory}, vol.~40, no.~3, pp. 721--731, May 1994.

\bibitem{BB84}
C.~H. Bennett and G.~Brassard, ``Quantum cryptography: Public key distribution
  and coin tossing,'' in \emph{Proc. of IEEE International Conference on
  Computers, Systems and Signal Processing}, Bangalore, India, Dec 1984, pp.
  175--179.

\bibitem{CoxHi}
D.~R. Cox and D.~V. Hinkley, \emph{Theoretical Statistics}.\hskip 1em plus
  0.5em minus 0.4em\relax London, {UK}: Chapman and Hall, 1974.

\end{thebibliography}

\end{document}